\documentclass{aa}[letter]  

\usepackage{graphicx}

\usepackage{txfonts}

\usepackage{xcolor}
\usepackage{graphicx}
\usepackage{orcidlink}
\usepackage{color, lastpage}
\usepackage{float}

\newcommand\newtext[1]{{#1 \bf}}

\usepackage{natbib}
\bibpunct{(}{)}{;}{a}{}{,} 

\begin{document}

\title{Asteroid modelling by starlight diffraction: \\The shape of Dimorphos, the satellite of (65803) Didymos}

\author{P. Tanga$^{\orcidlink{0000-0002-2718-997X}}$\inst{1},
K. Tsiganis$^{\orcidlink{0000-0003-3334-6190}}$\inst{2},
D. Souami$^{\orcidlink{0000-0003-4058-0815}}$\inst{3,4}\thanks{Fulbright Visiting Scholar (2022 - 2023) at University of California, Berkeley},
\newtext{R. Anderson}\inst{11},
E. Barbotin\inst{5}, 
A. Cazaux\inst{5},
F.~Colas$^{\orcidlink{0000-0002-0764-5042}}$\inst{7},
J. Hanu\v{s}$^{\orcidlink{0000-0002-2934-3723}}$\inst{8},
F. Marchis$^{\orcidlink{0000-0001-7016-7277}}$\inst{9,10}, 
J-L. Dauvergne\inst{5}, 
G. Langin\inst{5}, 
A. Leroy$^{\orcidlink{0009-0006-4865-6422}}$\inst{6}, 
B. Lott\inst{5},
A. Manna\inst{11},
L. Rousselot\inst{5}, 
A. Siakas\inst{2},
S. Sposetti\inst{11},
Ch. Vigna\inst{5},
F. Weber\inst{5},
A. W\"{u}nche\inst{5}   }
\institute{
Universit\'e C\^ote d’Azur, Observatoire de la C\^ote d’Azur, CNRS, Laboratoire Lagrange, Bd de l’Observatoire, CS 34229, 06304 Nice Cedex 4, France
 \and
 Aristotle University of Thessaloniki (AUTh), 54124, Greece
 \and
LIRA, CNRS UMR8254, Observatoire de Paris, Universit\'e PSL, Sorbonne Université, Université Paris Cité, CY Cergy Paris Université, Meudon, 92190, France
  \and
 naXys, Department of Mathematics, University of Namur, Rue de Bruxelles 61, 5000 Namur, Belgium
 \and
 Association Française d'Astronomie, 17, rue Emile Deutsch de la Meurthe, 74014 Paris, France
 \and
Uranoscope de l'Ile de France, All\'ee Camille Flammarion, 77220 Gretz-Armainvilliers, France
 \and
 LTE, Observatoire de Paris, Paris Sciences et Lettres University, 77 av. Denfert-Rochereau, 75014 Paris cedex, France
 \and
Charles University, Faculty of Mathematics and Physics, Institute of Astronomy, V Hole\v sovi\v ck\'ach 2, 18 000, Prague, Czech Republic\and
SETI Institute, Carl Sagan Center, 189 Bernado Avenue, Mountain View CA 94043, USA
\and
Unistellar, 5 allée Marcel Leclerc, bâtiment B, 13008 Marseille, France
 \and
International Occultation Timing Association/European Section,
Am Brombeerhag 13, 30459 Hannover, Germany
}

\date{}

  \abstract
   {The DART spacecraft impacted Dimorphos, the satellite of (65803) Didymos, in September 2022. Evidence of crater formation and possible global reshaping has been obtained indirectly from spacecraft and ground-based data.}
   {Since the impact, several stellar occultations by Didymos have been observed, but only one in particular, on January 21, 2023, can provide useful constraints on the size and shape of Dimorphos.}
   {We modelled the diffraction signatures recorded on multiple occultation chords to constrain the projected shape and size of Dimorphos, assuming an ellipsoidal model. This is the first time diffraction observed simultaneously on several chords of a single event has been used for such a purpose.}
   {The projected shape at the epoch of the event is well constrained and consistent with recent post-DART determinations. When extended to a full three-dimensional ellipsoidal solution, the result remains compatible with previous studies,  suggesting an equatorially elongated post-impact shape. }
   {}


\titlerunning{Diffraction modelling of the shape of Dimorphos}
\authorrunning{Tanga et al.}

   \maketitle
%

\section{Introduction}

The classical wave theory of light propagation describes the diffraction occurring at the edge of screens occulting a light source: a phenomenon that sometimes can also be observed when a Solar System object occults a distant star. For instance, in the past, modelling of the measured signal of diffraction during lunar occultation events has been exploited to measure the extension and limb darkening of large stellar discs \citep[][and references therein]{Richichi2011}. 

In the frame of studies of minor planets, the possible discovery of faint and distant cometary nuclei via the serendipitous measurement of a diffraction pattern, cast by a star, has been studied \citep{Roques87, roques}. \newtext{Detections of diffraction by Kuiper belt objects have been reported by \citet{Arimatsu2019, Schlichting2009}.} Occultations by main-belt or near-Earth asteroids (NEAs) often exhibit signal fluctuations at the ingress or egress of the event due to diffraction. Recently, diffraction measured on a single chord has been exploited to constrain the shape of the NEA (98943)~2001~CC21 by \citet{Arimatsu2024}. A general review including diffraction effects in stellar occultations has recently been published by \citet{Sicardy26}.

In this article, we investigate the shape and size of Dimorphos, the satellite of (65803)~Didymos hit by DART in September 2022, using a well-observed occultation event. Diffraction at the edge of the target, well visible in several light curves, allows us to constrain the shape and size of Dimorphos. The occultation took place on January 21, 2023, nearly four months after the impact. A crater due to the material ejection triggered by DART, or a deformation at larger scale, could have modified the shape. The aftermaths of the impact have been investigated in several articles indicating evidence of an overall oblate ellipsoid from DART images \citep[][]{Barnouin2023, Daly2024}, but possibly compatible with an elongated shape when other data are exploited \citep{Naidu24, Pravec2024, Zinzi24}, 

A single occultation event constrains only the two-dimensional projection of the target’s shape. However, by introducing physically motivated priors -- for example, plausible density or volume constraints -- it is possible to derive meaningful limits on the full three-dimensional, post-impact shape of Dimorphos.

In Section~\ref{s:occultprop} we describe the occultation conditions and the collected data. Their interpretation, exploiting a diffraction model, is presented in Sect.~\ref{s:datareduction}. The results of the processing provide constraints on the shape of Dimorphos that are discussed in Sect.~\ref{s:results}. In Sect.~\ref{s:conclusions} we present some more general implications of our findings. 

\section{Occultation properties and observations}
\label{s:occultprop}

The event that occurred on January 21, 2023 was one of the last of the observability season and was easily observable from Europe. We can estimate the Fresnel scale of diffraction as $L = \sqrt{ \lambda D / 2 }$, 
where $D$ is the distance from the occulting object at the event epoch, and $\lambda$ the wavelength. As is shown in Table~\ref{table:eventdata}, in this case $L$ is of  the same order as the size of Dimorphos. This is a very important feature for the interpretation of the recorded occultation chords, as we shall see further below.

For an observed placed on the Earth surface in the occultation path, the diffraction pattern moves along with the asteroid geometric shadow. The timescale associated with this variation can be written as

\begin{equation}
\tau_L = 0.20 \cdot \frac{10^3\,m\,s^{-1}}{v} \left[\, \left(\frac{\lambda}{550\ nm}\right) \left(\frac{D}{1\ au}\right)  \,\right]^\frac{1}{2} \,s
,\end{equation}

where $v$ is the apparent velocity of the asteroid shadow projected on a plane perpendicular to the direction of the star (Bessel, or fundamental, plane). For our specific event, the relatively slow shadow velocity and the bright target star are conditions that, combined, were favourable to a good sampling of the signal. 

Several observers, coordinated by the ACROSS network\footnote{https://www.oca.eu/fr/home-across} and the Association Française d'Astronomie, participated in the campaign. They are listed in Tab.~\ref{t:observers}. Three of them were not able to record data as they were clouded out or had technical problems. 

\section{Data reduction}
\label{s:datareduction}
\subsection{Fit to Didymos and timing calibration}

Before modelling Dimorphos, we first used the Didymos occultation to validate chord geometry and correct possible timing offsets. Given the short duration of the event, even sub-second timing errors would significantly bias the results.

The light curves were reduced following standard photometric techniques.
By projecting the chords on the normal plane $(f,g)$ \citep{Herald20, SORA}, we fitted the chords to the shape of Didymos. Since its diameter is $\sim$5.6 times larger than the Fresnel length, a simple knife-edge approximation for the occulting body was exploited. Diffraction was modelled independently for each light curve, and the epochs for the geometric disappearance and reappearance of the star were derived.

The observed chords (Tab.~\ref{t:observers}, \ref{t:observations} ) are labelled A to G (an index, $2$, was added for the same chords, referred to Dimorphos). Chords C and D have the best timing equipment, with no delays introduced by electronics, so they can be considered as very reliable. On the other hand, chords A and G had different technical issues potentially affecting the absolute timing.  

By adopting the pole solution by \citep{Scheirich2024}, we first projected on the fundamental plane two ellipsoidal shape models: the one of \citet{Barnouin2023} (818 × 796 × 590~m) and the one of \citet{Naidu24} (788 x 788 x 584~m). We immediately found that only the second one, producing a smaller projection, is compatible with our data.

\begin{figure}
\centering
\includegraphics[width=0.5\textwidth]{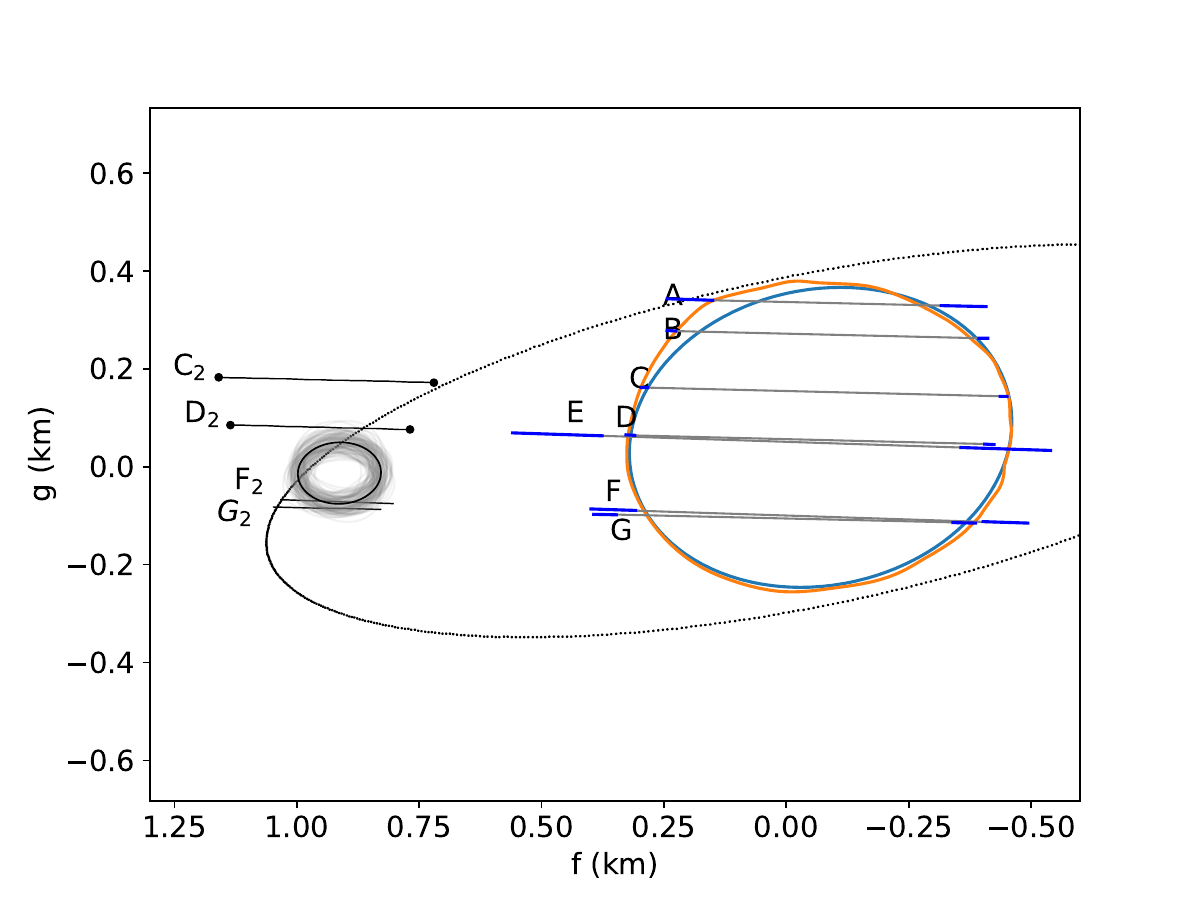}
\caption{Geometry of the occultation on the fundamental plane. The chords that produced positive detections are reported for Didymos, along with their error bars (grey segments). The two profiles for Didymos represent the ellipsoidal model by \citet{Naidu24} \newtext{(blue) and the topographic model by DRACO (orange)}. The chord segments to the left mark the approximate location and length of occultation features seen in the light curves in proximity of Dimorphos. The final profile reconstructed by diffraction Dimorphos is drawn. Light grey ellipses sample its errors on size, flattening, orientation, and position. The nominal orbit of Dimorphos follows the JPL s432 ephemeris.}
\label{occGeometry}
\end{figure}

We also tested the last post-DART topographic shape model (including DRACO images) available at the ESA kernel database\footnote{https://spiftp.esac.esa.int/data/SPICE/HERA/misc/hera.html}. 
We used the topographic shape to fit its projection to the observed chord extremes. The co-ordinates of the geometric centre of Didymos are the only free parameters. For the fit, we did not consider chords A and G for the issues mentioned above. The free parameters of our model are $f_0,g_0$, the co-ordinates of the shape centre on the plane. We obtain ($f_0,g_0$)$=(-0.082,0.012)$~km. Having derived the position of Didymos, we can adjust the position of chords A and G, which have timing errors, with respect to the shape, with time delays of -0.07~s and +0.565~s, respectively. The result of the procedure is shown in Fig.~\ref{occGeometry}.

\subsection{Diffraction modelling}

\begin{figure*}
\centering
\includegraphics[width=1.0\textwidth]{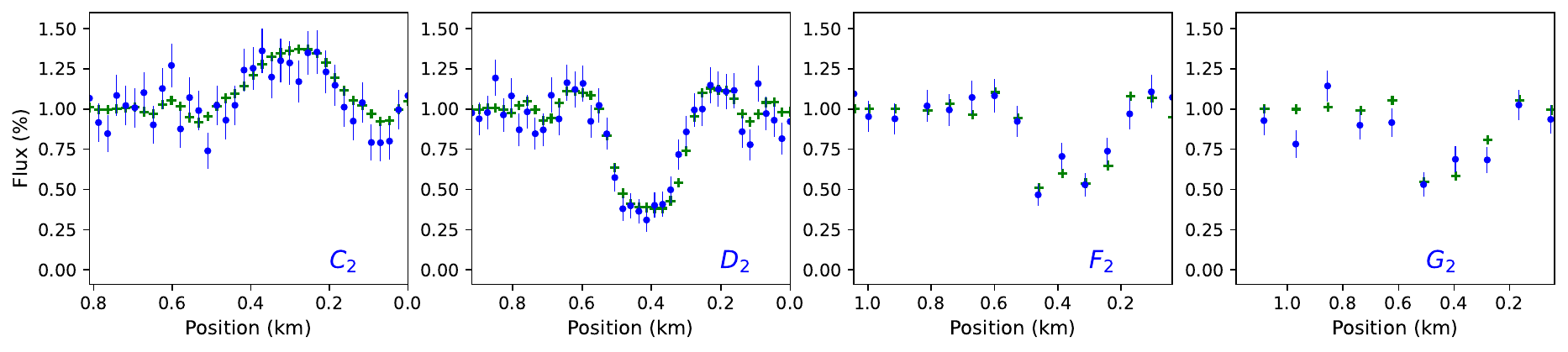}
\caption{Photometry of the occultation chords exploited to model the shape of Dimorphos. Note that none of the chords show a total extinction of light, and that local increases in the star brightness are also present. The blue dots represent the photometry and its error bars, while the green crosses indicate the best model fit (Sect.~\ref{s:results}).}
\label{f:LC}
\end{figure*}

This analysis is based on the occultation chords $C_2$, $D_2$, $F_2$, and $G_2$, (the suffix '2' indicates the chord portion related to Dimorphos), which show features related specifically to the satellite (Fig.~\ref{f:LC}). Chord $E_2$ (due to strong absorption) is too coarsely sampled to provide a detailed signal for the occultation by Dimorphos, so it is not considered in the following.

We also note that chords $F_2$ and $G_2$ have a lower time resolution. However, the fact that they are also very close to each other provides a mutual confirmation of the photometry and an advantage for cross-checking, especially since the chord F had a rather reliable time syncing thanks to Network Time Protocol.
Interestingly, we note that all the minima in chords $C_2$, $F_2$ and $G_2$ are suggestive of an occultation signal. However a na\"{i}ve interpretation assuming a geometric occultation would result in a very large size for Dimorphos. Moreover, several features due to diffraction appear: the brightening in chord $C_2$; the bright 'shoulders' on both sides of the minimum in $D_2$; and a possible central peak in $F_2$ and $G_2$.

\begin{figure}
\centering
\includegraphics[width=0.45\textwidth]{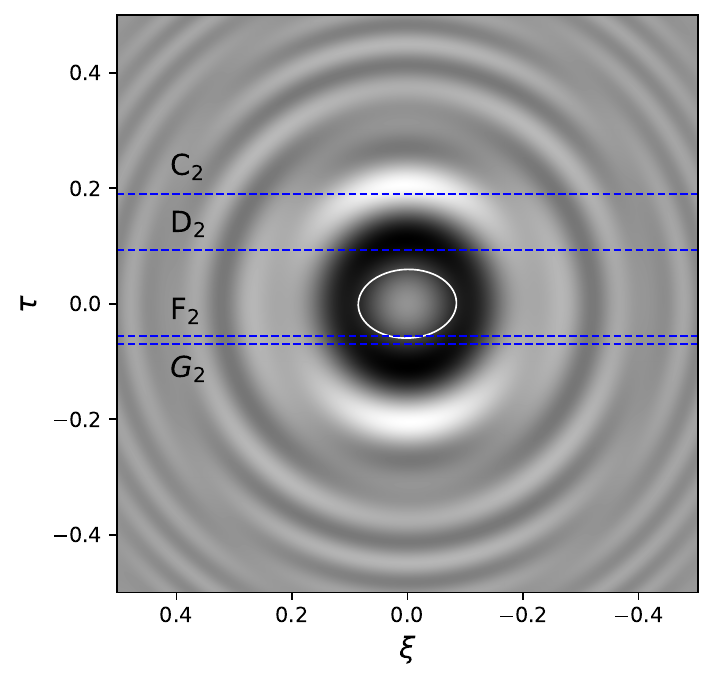}
\caption{Diffraction pattern created by Dimorphos, with the geometry of the chords (dotted lines). The horizontal and vertical axis are, respectively, the along- and cross- track directions. The white ellipse represents the geometric projection of the shape of Dimorphos. The background shows the diffraction pattern sampled by the four observers, with a large Poisson spot at the centre. Only one chord ($F_2$) is a geometric occultation; the one at the bottom ($G_2$) is just grazing.}
\label{diffShape}
\end{figure}

In order to model the signal, we considered an elliptical occulting screen representing the projected profile of Dimorphos. Because the Fresnel scale (L$\sim$140~m) is comparable to the size of Dimorphos, a one-dimensional knife-edge approximation is not valid. The full two-dimensional diffraction pattern of the projected shape must therefore be computed. This approach allows us to compute the shape whose diffraction pattern produces the simultaneous best fit to all the observed chords (Fig.~\ref{diffShape}).

Our model was initialised with the solution of \citet{Naidu24} for the post-DART ellipsoidal shape of Dimorphos. Its semi--axis dimensions, on the principal inertia axis, are (96 ± 6, 74 ± 4, 59 ± 7)~m. We call this shape the 'nominal ellipsoid'. Exploiting the fast Fourier transform, we computed the two-dimensional diffraction pattern \citep{Goodman1968,Trester1999} on a 2048$^2$ grid, centred on Dimorphos. Considering the typical frequency response of the cameras used for this campaign, we further numerically integrated the pattern over a range of wavelengths. The resulting flux map was then sampled along the observed chords, to obtain the model values of each light curve.

The vector of free parameters to determine is $\boldsymbol{V}(a,\Phi,\eta,f,g)$, in the order of: the ellipse major axis, the ellipse flattening, the position angle of the major axis, and the position ($f_d, g_d$) of the centre of the ellipse representing Dimorphos on the sky plane.
Starting from the nominal ellipsoid, we first executed a sequence of fits (with a Nelder-Mead minimisation algorithm) of the four chords concerning Dimorphos for: the size, orientation, and position of Dimorphos relative to the observed chords, effective wavelength, and passband. A unique position is compatible with the observation, with the centre \newtext{located between the chords $D_2$ and $F_2$}. For the effective wavelength $\lambda_{eff}$ and passband of the system we obtain $\lambda_{eff}$=530~nm and  $\Delta \lambda$ =130~nm for chords $C_2$ and $D_2$, and $\lambda_{eff}$=590~nm (same passband) for chords $F_2$ and $G_2$. At the noise level of our light curves, these values are not sensitive for changes of 10-20~nm in both quantities.

The fit quality is very sensitive to variations in the size and the position of Dimorphos at the level of $\sim$10~m only. These two quantities also appear rather uncorrelated. We exploited this property in the final processing step in which the state vector was split into two parts: $\boldsymbol{V'}(f_d,g_d)$ and $\boldsymbol{V''}(a,\Phi,\eta)$. 
$\boldsymbol{V'}$ was then sampled by the affine invariant Markov chain Monte Carlo algorithm of \citet{Goodman1968}, followed by the same procedure on $\boldsymbol{V''}$ (setting $\boldsymbol{V'}$ to its best solution). We then iterated the process, each time using the best parameters found at the previous step. Two iterations are sufficient, however, to show that the solution is very stable.

\section{Results}
\label{s:results}

The final result of the MCMC iteration (Fig.~\ref{2DshapeMCMC}) is represented in Fig.~\ref{occGeometry}. The projected shape of Didymos has dimensions $(a',b')\ =\ (170\pm10, 124\pm40)$~m, compatible, within the error bars, with a projection of a 'nominal ellipsoid'. This requires the sub-Earth point on Dimorphos to be situated at intermediate latitudes (assuming the pole direction along the 'c' axis). The maximum extension is also shortened by projection by a small amount, indicating that the 'a' axis of the nominal ellipsoid should not be far from parallel to the fundamental plane.  

The diffraction pattern relative to the four relevant chords and to the projected shape of Dimorphos are shown in Fig.~\ref{diffShape}. According to the value of $\eta$, the ellipse is elongated nearly exactly along the chords, with an error margin of $\sigma_\eta\sim13.5$ degrees. This orientation is critical to produce the correct brightness observed at the ‘shoulders’ of the minima, and the brightening in chord $D_2$. One can note that only chord $F_2$ has a geometric occultation. Diffraction, however, does not allow one to reach the complete stellar extinction. The increase in brightness at the bottom of the minimum corresponds to the large 'Poisson spot' at the centre. All the other chords bring signatures associated with features of the diffraction field, without being geometrically occulted. 

Concerning the position of Dimorphos, the formal accuracy from the MCMC fitting is $\sim$15~m, without a sensible difference between the cross-track and along-track directions. Considering that the displacement on the fundamental plane of Dimorphos relative to Didymos between the first and the last observation was about 20~m, we take this value as a better, conservative estimate for accuracy.
The resulting astrometry on the fundamental plane places Dimorphos at these relative distances from the centre of Didymos: $\Delta f = +1046 \pm 20$~m; $\ \Delta g = -104 \pm 20$~m.

The derived two-dimensional ellipse is compatible in principle with an infinite number of three-dimensional ellipsoids, with different orientations. However we can exploit some a priori information to derive final constraints on the shape of Dimorphos. Its ellipsoidal model is represented by the axis $a,b,c$, its projection by the sub-observer latitude and longitude ($\lambda_{so}$, $\beta_{so}$), and a rotation around the viewing direction to match the orientation of the projected ellipse ($\Psi$).

To constrain the ellipsoid, we further assumed that its volume is within a $30\%$ margin from \citet{Daly2024}, and the axis ratio $b/a \in [0.6,0.9]$. On this base, we again ran a MCMC optimisation, exploiting the covariance matrix associated with the elliptical projection as uncertainty on its parameters (Fig.~\ref{3DshapeMCMC}). A positive correlation appears between $a$ and $b$, and a negative one with $c$, but overall the distribution of the dimensions clearly favours the most probable values. Both the final volume-equivalent diameter, $D_{eq} = 152.4 \pm 6.8$~m, and the size ratios (a/b=1.35, b/c=1.47) are close to the post-impact shape by \citet{Naidu24}. 

The sub-solar point is unconstrained in the priors, so that symmetric peaks (four for the longitude and two for the latitude) appear in the a posteriori distribution, consistent with the symmetry of the shape model. Taking the smallest positive values, we obtain ($\lambda_{so}$, $\beta_{so}$)$\sim(30^\circ,50^\circ)$. Given the large uncertainty, this last value is compatible with the $\sim42^\circ$ latitude of the sub-Earth point for the JPL s432 ephemeris of Didymos. 

\section{Conclusions}
\label{s:conclusions}

The January 21, 2023 stellar occultation provides a rare opportunity to study Dimorphos in a diffraction-dominated regime in which the Fresnel scale is comparable to the size of the body. In this configuration, diffraction carries significant shape information that extends well beyond the simple geometric disappearance and reappearance times traditionally used in occultation analysis. For each of the observed chords, the recorded signal contains information about the size and orientation of Dimorphos.

For the first time, we simultaneously fit a two-dimensional diffraction pattern to multiple observed chords of a single event. This approach allows us to constrain not only the position of Dimorphos relative to Didymos, but also the size, orientation, and flattening of its projected ellipse. The solution is robust and highly sensitive to variations at the $\sim$10 m level, demonstrating the power of multi-chord diffraction modelling when the observational sampling and timing are sufficiently accurate.

The derived projected dimensions are fully consistent with post-DART determinations obtained from spacecraft imaging and radar measurements. When extending the solution to a three-dimensional ellipsoidal model, the intrinsic projection degeneracy becomes apparent. By introducing physically motivated priors on volume and axis ratios, we obtain a preferred solution compatible with previous studies, \newtext{supporting an elongated shape after the DART impact}. However, given the limitations of a single event, this three-dimensional inference should be regarded as indicative only.

Our analysis also highlights an important practical consequence: in diffraction-dominated occultations, measurable photometric signatures can occur well beyond the geometric limb of the object. As a result, the effective visibility track of very small asteroids may be significantly wider than predicted by geometric optics alone. In such cases, the traditional distinction between ‘positive’ and ‘negative’ chords becomes ambiguous; reporting and archiving standards may need to evolve accordingly.

Future occultations of Dimorphos and other small bodies observed in similar regimes will provide additional constraints and may help break the remaining degeneracies. As increasingly precise astrometry and coordinated multi-station observations become available, diffraction modelling will likely become an essential tool for characterising small bodies in the Solar System.

\begin{acknowledgements}
This work has been supported by ACROSS (Asteroid Collaborative Research via Occultation Systematic Surveys), and by the BQR programs (Observatoire de la C\^ote d'Azur and Lagrange Laboratory). The Czech Science Foundation has supported the research of JH (grant 22-17783S). We acknowledge the use of SORA~\citep{SORA}.
\end{acknowledgements}

\bibliographystyle{aa}
\bibliography{mybib}

\begin{thebibliography}{17}
\expandafter\ifx\csname natexlab\endcsname\relax\def\natexlab#1{#1}\fi

\bibitem[{Arimatsu {et~al.}(2019)Arimatsu, Tsumura, Usui, Shinnaka, Ichikawa, Ootsubo, Kotani, Wada, Nagase, \& Watanabe}]{Arimatsu2019}
Arimatsu, K., Tsumura, K., Usui, F., {et~al.} 2019, 1

\bibitem[{{Arimatsu} {et~al.}(2024){Arimatsu}, {Yoshida}, {Hayamizu}, {Ida}, {Hashimoto}, {Abe}, {Akitaya}, {Aratani}, {Fukuda}, {Fujita}, {Fujiwara}, {Horikawa}, {Iihoshi}, {Imamura}, {Imazawa}, {Kasebe}, {Kawasaki}, {Kishimoto}, {Mishima}, {Miyachi}, {Mizutani}, {Nakajima}, {Nakatani}, {Okamura}, {Okanobu}, {Okuda}, {Suzuki}, {Tatsumi}, {Uno}, {Yamamura}, {Yasue}, {Yoshihara}, {Hirabayashi}, \& {Yoshikawa}}]{Arimatsu2024}
{Arimatsu}, K., {Yoshida}, F., {Hayamizu}, T., {et~al.} 2024, \pasj, 76, 940

\bibitem[{Barnouin {et~al.}(2023)Barnouin, Ballouz, Marchi, Vincent, Pajola, Luchetti, Daly, Ernst, Palmer, Gaskell, Kahout, Robin, Murdoch, Sunshine, Farnham, Tusberti, Rizos, Zhang, Ferrari, Agrusa, Hirabayashi, Parro, Cambioni, Michel, Ruducan, Jutzi, Asphaug, Nolan, Campo~Bagatin, Trigo-Rodríguez, Zinzi, Della~Corte, Chabot, Rivkin, Cheng, Dotto, Team, \& Team}]{Barnouin2023}
Barnouin, O.~S., Ballouz, R.~L., Marchi, S., {et~al.} 2023, 54th LPSC, 2806, 2605

\bibitem[{{Daly} {et~al.}(2024){Daly}, {Ernst}, {Barnouin}, {Gaskell}, {Nair}, {Agrusa}, {Chabot}, {Cheng}, {Dotto}, {Mazzotta Epifani}, {Espiritu}, {Farnham}, {Palmer}, {Pravec}, {Rivkin}, {Waller}, {Zinzi}, {DART Team}, \& {LICIACube Team}}]{Daly2024}
{Daly}, R.~T., {Ernst}, C.~M., {Barnouin}, O.~S., {et~al.} 2024, Planeary Science Journal, 5, 24

\bibitem[{{Gomes-J{\'u}nior} {et~al.}(2022){Gomes-J{\'u}nior}, {Morgado}, {Benedetti-Rossi}, {Boufleur}, {Rommel}, {Banda-Huarca}, {Kilic}, {Braga-Ribas}, \& {Sicardy}}]{SORA}
{Gomes-J{\'u}nior}, A.~R., {Morgado}, B.~E., {Benedetti-Rossi}, G., {et~al.} 2022, \mnras, 511, 1167

\bibitem[{Goodman(1968)}]{Goodman1968}
Goodman, J.~W. 1968, Introduction to Fourier optics | BibSonomy (New York, NY: McGraw-Hill)

\bibitem[{Herald {et~al.}(2020)Herald, Gault, Anderson, Dunham, Frappa, Hayamizu, Kerr, Miyashita, Moore, Pavlov, Preston, Talbot, \& Timerson}]{Herald20}
Herald, D., Gault, D., Anderson, R., {et~al.} 2020, MNRAS, 499, 4570–4590

\bibitem[{Naidu {et~al.}(2024)Naidu, Chesley, Moskovitz, Thomas, Meyer, Pravec, Scheirich, Farnocchia, Scheeres, Brozovic, Benner, Rivkin, \& Chabot}]{Naidu24}
Naidu, S.~P., Chesley, S.~R., Moskovitz, N., {et~al.} 2024, The Planetary Sc. Journal, Volume 5, Issue 3, id.74, 11 pp., 5, 74

\bibitem[{Pravec {et~al.}(2024)Pravec, Meyer, Scheirich, Scheeres, Benson, \& Agrusa}]{Pravec2024}
Pravec, P., Meyer, A.~J., Scheirich, P., {et~al.} 2024, Icarus, 418, 116138

\bibitem[{{Richichi} {et~al.}(2011){Richichi}, {Chen}, {Fors}, \& {Wang}}]{Richichi2011}
{Richichi}, A., {Chen}, W.~P., {Fors}, O., \& {Wang}, P.~F. 2011, \aap, 532, A101

\bibitem[{{Roques} {et~al.}(2008){Roques}, {Georgevits}, \& {Doressoundiram}}]{roques}
{Roques}, F., {Georgevits}, G., \& {Doressoundiram}, A. 2008, {The Kuiper Belt Explored by Serendipitous Stellar Occultations} ({Barucci}, M.~A. and {Boehnhardt}, H. and {Cruikshank}, D.~P. and {Morbidelli}, A. and {Dotson}, R.), 545--556

\bibitem[{{Roques} {et~al.}(1987){Roques}, {Moncuquet}, \& {Sicardy}}]{Roques87}
{Roques}, F., {Moncuquet}, M., \& {Sicardy}, B. 1987, \aj, 93, 1549

\bibitem[{Scheirich {et~al.}(2024)Scheirich, Pravec, Meyer, Agrusa, Richardson, Chesley, Naidu, Thomas, \& Moskovitz}]{Scheirich2024}
Scheirich, P., Pravec, P., Meyer, A.~J., {et~al.} 2024, The Planetary Sc. Journal, 5, 17

\bibitem[{Schlichting {et~al.}(2009)Schlichting, Ofek, Wenz, Sari, Gal-Yam, Livio, Nelan, \& Zucker}]{Schlichting2009}
Schlichting, H.~E., Ofek, E.~O., Wenz, M., {et~al.} 2009, Nature, 462, 895–897

\bibitem[{{Sicardy} \& {Dettwiller}(2026)}]{Sicardy26}
{Sicardy}, B. \& {Dettwiller}, L. 2026, \aap, 707, A208

\bibitem[{Trester(1999)}]{Trester1999}
Trester, S. 1999, Computing in Science and Engineering, 1, 77–83

\bibitem[{Zinzi {et~al.}(2024)Zinzi, Hasselmann, Della~Corte, Deshapriya, Gai, Lucchetti, Pajola, Rossi, Dotto, Mazzotta~Epifani, Daly, Hirabayashi, Farnham, Ernst, Ivanovski, Li, Parro, Amoroso, Beccarelli, Bertini, Brucato, Capannolo, Caporali, Ceresoli, Cremonese, Dall’Ora, Gomez~Casajus, Gramigna, Ieva, Impresario, Lasagni~Manghi, Lavagna, Lombardo, Modenini, Negri, Palumbo, Perna, Pirrotta, Poggiali, Tortora, Tusberti, Zannoni, \& Zanotti}]{Zinzi24}
Zinzi, A., Hasselmann, P. H.~A., Della~Corte, V., {et~al.} 2024, The Planetary Sc. Journal, 5, 103

\end{thebibliography}

\begin{appendix}
\onecolumn

\section{Complementary tables and plots.}\label{ssec:appendixA}

\begin{table}[H]
\caption{Main properties of the occultation event.}             
\label{table:eventdata}
\centering                          
\begin{tabular}{l r}        
Closest geocentric approach to the star & 2023-01-21 23:33:17.620 UT\\
\\
Star & \\
\hline 
Gaia-EDR3 source ID & 886931023166371456\\
Magnitudes & G:  9.071, B:  9.260, V:  9.077, R:  8.970\\
Apparent diameter of the star (V) & 0.0650 mas Kervella et. al (2004)\\
\\
Asteroid & \\
\hline
Geocentric asteroid distance &     0.38 au\\
Geocentric shadow velocity & 1.965 km / s \\
\\
Diffraction scales & \\
\hline 
Fresnel scale L   &   0.14 km\\
Fresnel time $\tau_L$ & 0.072 s \\
\hline                                   
\end{tabular}
\end{table}
\begin{table}[H]
\caption{Observers and co-ordinates of thee observing sites.  }            
\label{t:observers}
\centering                          
\begin{tabular}{lrrrlrrl}
\hline
Observer	&    Long.	&	Lat. &  Alt. &  Instr.    & Sampling & Label & Comment \\
            &    (deg)  &   (deg)  &   (m)        &         &  (ms)       &                  &   \\
\hline
\hline
E. Barbotin - F. Colas& 45.210965 &-0.367708 & 62  & 1 & 25 & A & 1 / 5 frame recorded.  \\
J.-L. Dauvergne & 45.211667 &-0.501200 & 43  & 2 & 19 & B &  \\
A. Leroy    & 45.2092547&-0.365796& 62   & 3 & 50 & C & \\
L. Rousselot  & 45.1931666&0.8300556& 95   & 2 & 100 & D & \\
S. Sposetti - A. Manna& 44.983022  &7.480575& 283  & 4 & 80 & E & Heavy fogs.\\
J. Hanu\v{s}   & 44.996611 &7.133285 & 826  & 50 & 5 & F &  \\
B. Lott    & 45.2126217&-1.070888& 17   & 6 & 50 & G & Timing errors$\sim$0.5 s\\
G. Langin  &45.214372&-0  .502422& 44   & 7 & &   &   \\
F. Weber	  &  45.201956& 0.661028& 105  & 8 & &   &   \\ 
A. Cazaux  &45.19266	&     1.1344& 336  & 9 & &   &  No data.     \\   
A. W{\"u}nsche    &45.0546944&  5.793694& 436 & 10& &   & No data. \\
\hline

\end{tabular}
\tablefoot{
For observer each, longitude, latitude and elevation are provided. The labels of the occultation chords are given as used in the text of the article. \newtext{The sampling interval is provided; its consistency with the exposure time has been verified.}\\
Instrument codes: (1) 22.5 cm Schmidt-Cass., CCD+GPS; (2) 25 cm Schmidt-Cass., CMOS+Timebox; (3) 25 cm Schmidt-Cass., CCD+GPS; (4) 20 cm Schmidt-Cass., video + GPS; (5) Unistellar eVscope, NTP sync by phone; (6) 20 cm Schmidt-Cass., CMOS+Timebox; (7) 15 cm Newt., CMOS+Timebox; (8) 9 cm Refr. + SRL camera; (9) 18 cm Schmidt-Cass.; (10)  Unistellar eQuinox.}
\end{table}
\begin{table}[H]
\caption{Chords with positive detections.  }             
\label{t:observations}
\centering                          
\begin{tabular}{l l l l l}
\hline
Chord label	&   Didymos D   &      Didymos R      &         Dimorphos D      &     Dimorphos R \\  
\hline
\hline
G, G$_2$          &  23:30:48.68  ± 0.01   &  23:30:49.00   ± 0.01  &  23:30:49.22  ± 0.01 & 23:30:49.29  ± 0.01\\
F, F$_2$      &  23:26:08.64  ± 0.02   &  23:26:08.99   ± 0.02  &  23:26:09.22  ± 0.02 & 23:26:09.30  ± 0.02 \\
A &  23:30:25.262 ± 0.02   &  23:30:25.506  ± 0.02 &&\\
C, C$_2$        &  23:30:25.087 ± 0.003  &  23:30:25.407  ± 0.003 &  23:30:25.69  ± 0.01 & 23:30:25.71  ± 0.01\\
D, D$_2$     &  23:29:44.186 ± 0.006  &  23:29:44.506  ± 0.006 &  23:29:44.732 ± 0.006& 23:29:44.833 ± 0.006 \\
B     &  23:30:29.771 ± 0.004  &  23:30:30.049  ± 0.004 & & \\
E &  23:25:56.81  ± 0.05   &  23:25:57.17   ± 0.05   & & \\

\hline
\end{tabular}
\tablefoot{
Epochs are UT of January 21, 2023. For Dimorphos, D and R times correspond to the beginning and end of light--curve features as the geometric occultation is not observed directly.
}
\end{table}

\begin{figure}[H]
\centering
\includegraphics[width=0.7\textwidth]{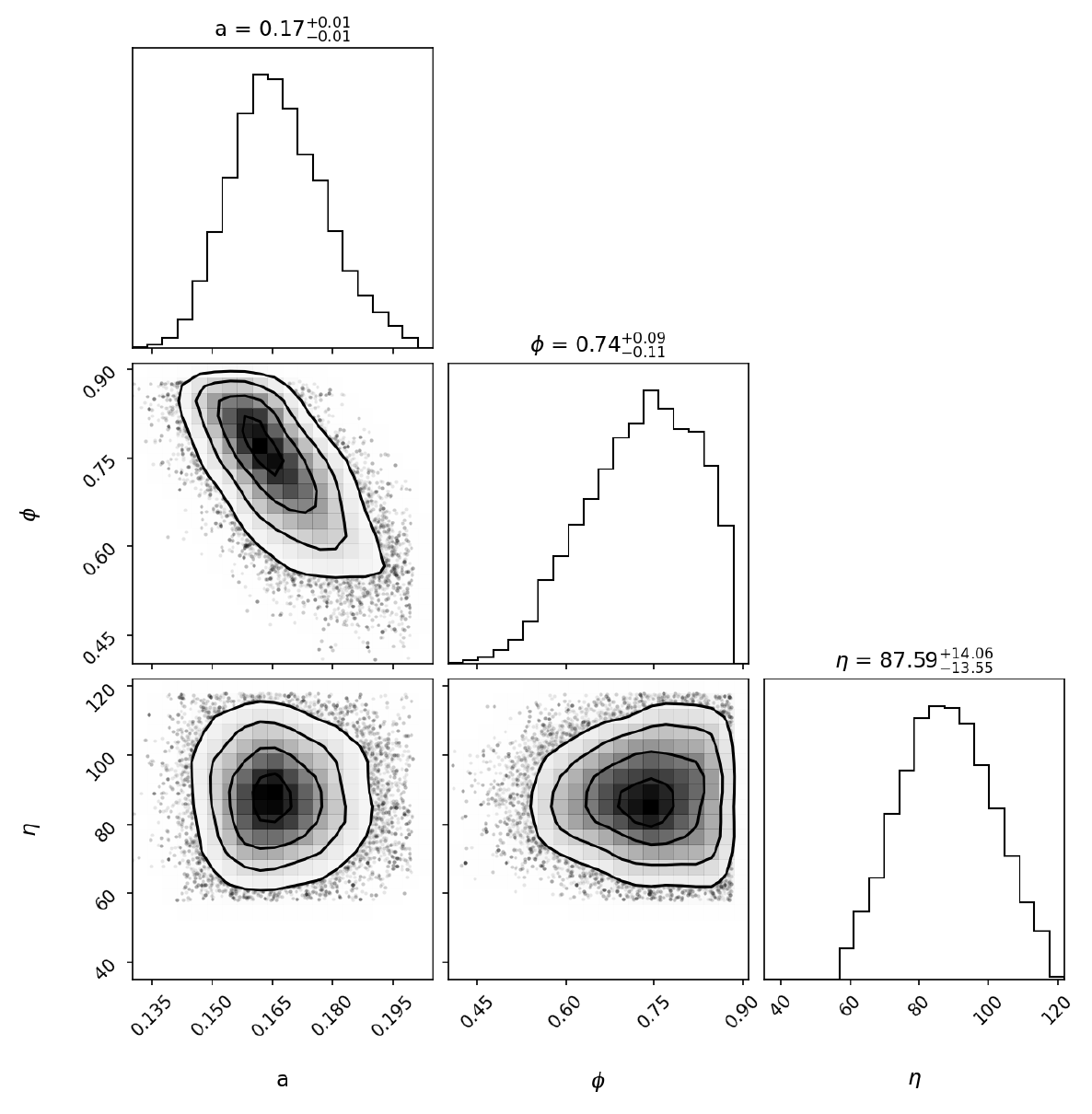}
\caption{Triangular plot of the MCMC results for the determination of $\boldsymbol{V''}(a,\Phi,\eta)$ as described in the text. The parameters are the
ellipse major axis length (m), its flattening and the position angle of the
major axis (degrees). A correlation can be observed between the size, $a$, and the flattening, only. The peak value and uncertainties are provided on the top of each column.}
\label{2DshapeMCMC}
\end{figure}
\clearpage
\begin{figure}
\centering
\includegraphics[width=0.75\textwidth]{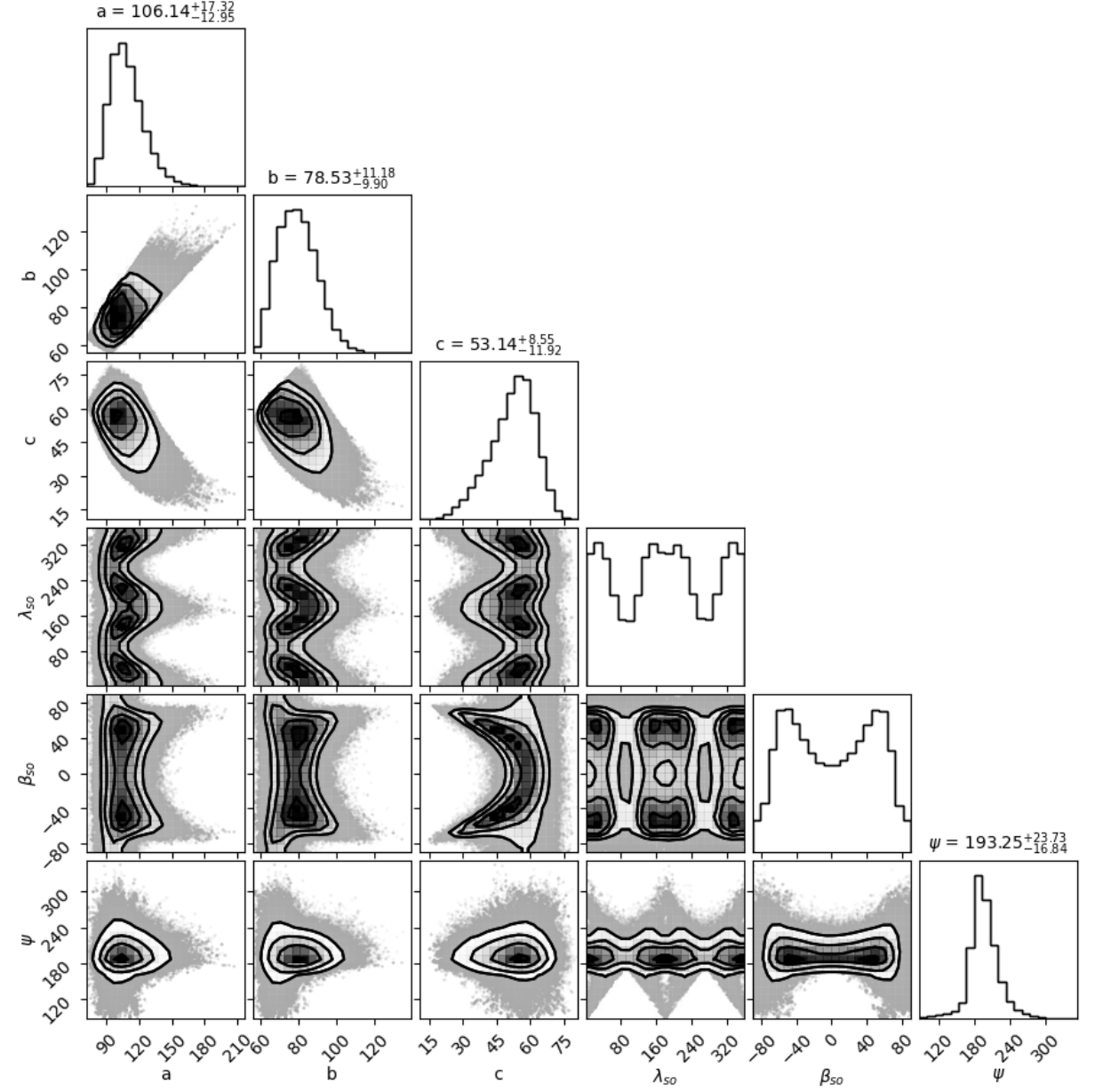}
\caption{Results of the final MCMC run, used to sample the possible shape and orientation of an ellipsoidal Dimorphos, with the constraints obtained on its projection at the occultation epoch (Fig.~\ref{2DshapeMCMC}). The ellipse semi--axis, the longitude and latitude of the sub-Earth point, and the position angle of the N pole on the sky are represented. The ellipsoid axis and the position angle are reasonably constrained. The sub-Earth point approximate longitude and latitude, for symmetry reasons, have a double peak distribution. We define the latitude as the angle at the ellipsoid centre with respect to the $(a,b)$ plane. The longitude origin is taken in the direction of the longest axis, $a$. The origin of the rotation $\Psi$ is along the major axis of the projected ellipse, at position angle $\eta=87^\circ.6$.}
\label{3DshapeMCMC}
\end{figure}

\end{appendix}

\end{document}